\documentclass[prl,aps,reprint,a4paper,superscriptaddress,floatfix,longbibliography]{revtex4-2}
\usepackage{times}
\usepackage{graphicx}
\usepackage{epsfig}
\usepackage{amsfonts}
\usepackage{amsmath}
\usepackage{amssymb}
\usepackage{dsfont}
\usepackage{amsthm}% in order to use a black box at the end of proofs
\usepackage{color}
\usepackage[colorlinks=true,linkcolor=blue,citecolor=blue,urlcolor=blue]{hyperref}
\usepackage{comment}
\usepackage{braket}
\usepackage{physics}
\usepackage{multirow}
\usepackage{diagbox}

\newcommand{\id}{\mathds{1}}

\begin{document}

%%%%%%%%%%%%%%%%%%%%%%%%%%%%%%%%%%%%%%%%%%%%%%%%%%%%%%%%%%%%%%%%%%%

\title{Simplest Bipartite Perfect Quantum Strategies}

%%%%%%%%%%%%%%%%%%%%%%%%%%%%%%%%%%%%%%%%%%%%%%%%%%%%%%%%%%%%%%%%%%%

\author{Ad\'an~Cabello}
\email{adan@us.es}
\affiliation{Departamento de F\'{\i}sica Aplicada II, Universidad de Sevilla, E-41012 Sevilla,
Spain}
\affiliation{Instituto Carlos~I de F\'{\i}sica Te\'orica y Computacional, Universidad de
Sevilla, E-41012 Sevilla, Spain}

%%%%%%%%%%%%%%%%%%%%%%%%%%%%%%%%%%%%%%%%%%%%%%%%%%%%%%%%%%%%%%%%%%%

\begin{abstract}
A bipartite perfect quantum strategy (BPQS) allows two players who cannot communicate with each other to always win a nonlocal game. BPQSs are rare but fundamental in light of some recent results in quantum information, computation, and foundations. A more than 40-year-old open problem is how many inputs (measurement settings) a BPQS requires. A related problem is how many inputs are needed if, in addition, the quantum system has minimum dimension. A third, apparently unrelated, problem is what is the connection between BPQSs and state-independent contextuality, which inspired the first BPQSs. Here, we solve the third problem: we prove that {\em every} BPQS defines a Kochen-Specker set. We use this result to identify the BPQS with the smallest number of inputs, both in the general case and in the case of minimum dimension, and solve some related problems. We conjecture that the BPQSs presented here are the solutions to the first two problems. 
\end{abstract}

%%%%%%%%%%%%%%%%%%%%%%%%%%%%%%%%%%%%%%%%%%%%%%%%%%%%%%%%%%%%%%%%%%%

\maketitle

%%%%%%%%%%%%%%%%%%%%%%%%%%%%%%%%%%%%%%%%%%%%%%%%%%%%%%%%%%%%%%%%%%%

{\em Introduction---}Bell nonlocality \cite{Bell:1964PHY,Brunner:2014RMP,Scarani2019} is arguably the clearest manifestation of quantum versus classical advantage. However, not all forms of Bell nonlocality are equally powerful and informative about how nature works. One way to realize this is to turn the violation of a Bell inequality into a nonlocal game in which two parties, Alice and Bob, who cannot communicate with each other, use a quantum correlation to win more often than by using the best classical strategy. In every trial, the referee gives Alice an input $x \in X$, and gives Bob an input $y \in Y$, with probability $\pi(x,y)$.
In return, Alice gives the referee an output $a \in A$, and Bob gives an output $b \in B$. Alice and Bob win the game if the inputs and outputs satisfy a certain condition $W(x,y,a,b) \in \{0,1\}$. A nonlocal game has a {\em bipartite perfect quantum strategy} (BPQS) if, using a quantum correlation $p(a,b|x,y)$, the probability of winning, $\omega = \sum_{x,y,a,b} \pi(x,y)\,p(a,b|x,y)\,W(x,y,a,b)$, is $1$ and no classical strategy allows them to always win the game. That is, there is a BPQS if
\begin{equation}
\omega_C < \omega_Q = 1.
\end{equation}

Although BPQSs have been known since 1983 \cite{Stairs:1983PS,HR83} and some of them (e.g., the ``magic-square'' correlation \cite{Cabello:2001PRLb}) are used frequently in quantum computation and quantum information, it is only recently that BPQSs have attracted increasing attention. Arguably, this is due to several results, including the following. First, it has been proven \cite{Li:2023XXX} that $p(a,b|x,y)$ is a BPQS if and only if (i) it has local fraction \cite{Elitzur:1992PLA} $0$ or is ``fully nonlocal'' \cite{Aolita:2012PRL,Aolita:2012PRA} or has ``strong contextuality'' \cite{Ramanathan:PRL2012,Abramsky:PRL2017},
(ii)~it allows for a Greenberger-Horne-Zeilinger-like \cite{GHZ89} or ``all-versus-nothing'' \cite{Cabello:2001PRLb,CinelliPRL2005,YangPRL2005} proof of nonlocality, 
or (iii)~it is in a face of the nonsignaling polytope that does not contain local points \cite{Goh:2018PRA}. 

Second, it has been proven that quantum theory cannot be explained by most {\em nonlocal} hidden-variable theories \cite{Vieira;2024XXX}. Remarkably, a necessary condition for $p(a,b|x,y)$ to violate a Bell-like inequality for nonlocal hidden-variable theories with arbitrary joint relaxations of the assumptions (made in Bell's theorem \cite{Bell:1964PHY}) of measurement independence and parameter independence \cite{Shimony:1993} is that $p(a,b|x,y)$ is a BPQS or $\epsilon$-close to a BPQS.

Third, there is an infinite-dimensional commuting BPQS for a nonlocal game whose maximum probability of winning with tensor product correlations is $\omega < \frac{1}{2}$ \cite{Ji:2021CACM}. This shows that there are quantum correlations produced by commuting measurements (which are the ones used in quantum field theory for describing correlations in Bell scenarios), which cannot be approximated by sequences of finite-dimensional tensor product correlations (which are the ones used in quantum information) \cite{Ji:2021CACM} and has deep consequences in physics \cite{cabello2023logicalpossibilitiesphysicsmipre}. 

Fourth, perfect quantum strategies allow us to explain the limits of all other quantum strategies \cite{Cabello:2013PRL, Cabello:2015PRL,Cabello:2019PRA}.

Fifth, multipartite perfect quantum strategies can be mapped into the circuit model of quantum computation so that the quantum advantage in the nonlocal games translates into certain problems being solvable by a quantum algorithm with certainty and in constant time for any input size, but requiring a time logarithmic in the length of the input for any classical circuit that solves them with a sufficiently high probability \cite{Bravyi:2018SCI}. This is the first proof of nonoracular quantum computational advantage.

Sixth, there are several quantum information protocols that require BPQSs \cite{Cubitt:2010PRL,Horodecki:2010XXX,Vidick:2017XXX,Jain:2020IEE,Zhen:2023PRL,Bharti:2023XXX}.

In short, BPQSs play crucial roles more and more frequently in quantum computation, information, and foundations. It is therefore surprising that, 40 years after their discovery, we do not have the answer to any of the following three problems. 

We know that there is no BPQS if $|X|=|Y|=3$, $|A|=4$, $|B|=2$, or any subset of it \cite{Li:2023XXX}. The BPQS with the smallest $|X| \times |Y|$ {\em known} has $|X|=|Y|=3$ and $|A|=|B|=4$ \cite{Cabello:2001PRLb}.

{\em Problem 1---}Which is the BPQS with the minimum number of inputs and outputs \cite{Stairs:1983PS,HR83,Gisin:2007IJQI}, that is, with minimum $|X| \times |Y|$ (in first place) and minimum $|A| \times |B|$ (in second place)?

BPQSs are impossible with a pair of qubits. A BPQS requires, at least, a qutrit-qutrit system \cite{Renner2004b,Brassard:2005}. The qutrit-qutrit BPQS with the smallest $|X| \times |Y|$ that can be found in the literature has $|X|=31$, $|Y|=17$, $|A|=2$, and $|B|=3$, and is obtained by combining the method in \cite{Elby:1992PLA,Renner2004b} with Conway and Kochen's 31-projector Kochen-Specker (KS) set \cite{Peres:1993}.

{\em Problem 2---}Which is the BPQS of minimum dimension (in first place) and minimum number of inputs in second place) and outputs (in third place) \cite{Stairs:1983PS,HR83,Gisin:2007IJQI}?

In 1983, Stairs \cite{Stairs:1983PS} and Heywood and Redhead \cite{HR83} showed that, if Alice and Bob share a maximally entangled state between two qudits of $d \ge 3$, and each of them measures the observables used for proving the KS theorem of impossibility of hidden-variable theories \cite{Kochen:1967JMM}, now called a KS set (defined below), the resulting correlation is what we now call a BPQS. This was the first example of a BPQS. For years, KS sets were also the only way to produce experimentally testable quantum state-independent contextuality (SI-C) \cite{Cabello:2008PRL,Badziag:2009PRL,Kirchmair:2009NAT,Amselem:2009PRL}. However, Yu and Oh \cite{Yu:2012PRL} showed that SI-C can also be produced by observables that do not form a KS set \cite{Yu:2012PRL,Bengtsson:2012PLA,Xu:2015PLA}. Recently, these SI-C (but not KS) sets have been used to produce quantum correlations \cite{Huang:2021PRL,Junior:2023PRR}. But can a SI-C set that is not a KS set produce a BPQS? Do we even need a SI-C set for BPQS? More generally,

{\em Problem 3---}What is the relation between BPQSs and SI-C?

Problems~1 and~2 cannot be solved with existing tools because the exhaustive exploration of the relation between the set of quantum correlations and the nonsignaling polytope for bipartite Bell scenarios beyond $|X|=|Y|=3$, $|A|=4$, $|B|=2$ is not computationally feasible \cite{Li:2023XXX}. Faced with the severe risk that Problems~1 and~2 remain open for another 40 years, and urged by the interdisciplinary importance that BPQSs have recently acquired, we propose a different approach. 

At first sight, Problem~3 is unconnected to the other two problems. 
However, as we will see, the solution Problem~3 not only forces us to change our view on the role of KS sets in quantum theory, quantum information, and quantum computation, it also gives us a new tool to attack Problems~1 and~2. On the one hand, it allows us to
prove that BPQSs are impossible in $|X|=|Y|=|A|=|B|=3$ and leads us to conjecture the solution to Problem~1. On the other hand, it allows us to reduce the number of inputs needed for a qutrit-qutrit BPQS and conjecture the solution to Problem~2. In addition, it solves or gives us clues to solve some other problems.

%%%%%%%%%%%%%%%%%%%%%%%%%%%%%%%%%%%%%%%%%%%%%%%%%%%%%%%%%%%%%%%%%%%

{\em Every BPQS defines a KS set---}Our first result is that {\em every} BPQS defines a special type of KS set. For detailing the result, we need two definitions.

{\em Definition 1 (Generalized KS set \cite{Xu:2024PRL})---}A generalized KS set is a set $\mathcal{V}$ of projectors, not necessarily of rank one, in a Hilbert space of finite dimension $d \ge 3$, which does not admit an assignment $f: \mathcal{V} \rightarrow \{0,1\}$ satisfying (i) For two orthogonal projectors $u, v \in \mathcal{V}$, $f(u) + f(v) \leq 1$, and (ii) for every subset $b$ of mutually orthogonal elements of $\mathcal{V}$ summing the identity, $\sum_{u \in b} f(u) = 1$.

Replacing ``not necessarily of rank one'' by ``of rank one,'' one recovers the standard definition of KS set \cite{Kochen:1967JMM,Renner2004b,Pavicic:2005JPA}.

{\em Definition 2 (Bipartite KS set)---}Let $S_\mathcal{A}$ ($S_\mathcal{B}$) be a set in which each element is a set of mutually orthogonal projectors summing to the identity, and let $\mathcal{V}_\mathcal{A}$ ($\mathcal{V}_\mathcal{B}$) be the set of all projectors in $S_\mathcal{A}$ ($S_\mathcal{B}$).
Two sets $S_\mathcal{A}$ and $S_\mathcal{B}$ constitute a bipartite KS set (B-KS set) if there is no assignment $f : (\mathcal{V}_\mathcal{A} \cup \mathcal{V}_\mathcal{B}) \times (S_\mathcal{A} \cup S_\mathcal{B}) \rightarrow \{0,1\}$ satisfying
(I') $f(u,b) + f(u',b') \leq 1$ for each $b \in S_\mathcal{A}, b' \in S_\mathcal{B}, u \in b, u' \in b'$ with $u,u'$ orthogonal,
(II') $\sum_{u \in b} f(u,b) = 1$ for each $b \in S_\mathcal{A}$, and $\sum_{u' \in b'} f(u',b') = 1$ for each $b' \in S_\mathcal{B}$.

Clearly, a B-KS set is a generalized KS set. 

Now recall that any quantum correlation $p(a,b|x,y)$ in a finite dimensional Hilbert space ${\mathcal H}={\mathcal H}_{\mathcal A} \otimes {\mathcal H}_{\mathcal B}$ can be realized as
\begin{equation}
\label{gene}
p(a,b|x,y) = \langle \psi | \Pi_{a|x} \otimes \Pi_{b|y} | \psi \rangle,
\end{equation}
where $\ket{\psi}$ is a pure state, Alice's measurements $\Pi_{a|x}$ are projective, and Bob's measurements $\Pi_{b|y}$ are also projective (see, e.g., \cite{Brunner:2014RMP}[Sec.\ II.A.3] and \cite{Paddock:2022XXX}). The reason is, on the one hand, that a simple purification argument shows that any correlation achieved with a mixed state can be achieved with a pure state \cite{Paddock:2022XXX,Sikora:2016PRL}. On the other hand, 
Naimark's dilation theorem \cite{Neumark:1940IANa} implies that any local measurement that can be achieved by a positive operator-valued measure (the most general way to represent a measurement in quantum theory) on a finite-dimensional Hilbert space can also be achieved by a projective-value measure on a larger finite-dimensional Hilbert space. 

We also need to recall that, according to quantum theory, any projective-value measure is, in principle, implementable in such a way that the quantum postmeasurement states satisfy L\"uder's rule \cite{Luders:1951APL}. That is, when Alice measures $\Pi_{a|x}$ and obtains $a$, and Bob measures $\Pi_{b|y}$ and obtains $b$, the initial state transforms as 
\begin{equation}
\label{l2}
|\psi\rangle \longrightarrow |\psi_{a,b|x,y}\rangle = \frac{(\Pi_{a|x} \otimes \Pi_{b|y}) |\psi\rangle}{\sqrt{\langle \psi |(\Pi_{a|x} \otimes \Pi_{b|y}) | \psi \rangle}}.
\end{equation}
Now we can present our result.

{\em Theorem 1---}Every BPQS $p(a,b|x,y)$ in a finite dimensional Hilbert space ${\mathcal H}={\mathcal H}_{\mathcal A} \otimes {\mathcal H}_{\mathcal B}$ such that $\dim({\mathcal H}_{\mathcal A}) \le \dim({\mathcal H}_{\mathcal B})$, produces a B-KS $(S_{\mathcal A}, S_{\mathcal B})$ using the following method.
First, find a realization of $p(a,b|x,y)$ with a pure state $|\psi\rangle$ and projective measurements $\Pi_{a|x}$ and $\Pi_{b|y}$ satisfying L\"uder's rule. Then, define $S_{\mathcal A}$ and $S_{\mathcal B}$ as follows:

$S_{\mathcal A}=\{s_x\}_{x \in X}$, where $s_x= \{\rho^{{\mathcal A}}_{a|x}\}_{a \in A}$
is the set of Alice's postmeasurement states for $x$. That is, 
\begin{equation}
\rho^{{\mathcal A}}_{a|x} = \tr_{\mathcal B}(|\psi_{a|x}\rangle \langle\psi_{a|x}|),
\end{equation}
where $\tr_{\mathcal B}$ denotes partial trace over ${\mathcal H}_\mathcal{B}$ and
\begin{equation}
\label{lal}
|\psi_{a|x}\rangle = \frac{(\Pi_{a|x} \otimes \id_{\mathcal B}) |\psi\rangle}{\sqrt{\langle \psi |(\Pi_{a|x} \otimes \id_{\mathcal B}) | \psi \rangle}},
\end{equation}
where $\id_{\mathcal B}$ is the identity in ${\mathcal H}_{\mathcal B}$. 

$S_{\mathcal B}=\{s_y\}_{y \in Y}$, where $s_y= \{\rho^{{\mathcal A}}_{b|y}\}_{b \in B}$ 
is the set of Alice's premeasurement states after Bob's measurement $y$. That is,
\begin{equation}
\rho^{{\mathcal A}}_{b|y} = \tr_{\mathcal B}(|\psi_{b|y}\rangle \langle\psi_{b|y}|),
\end{equation}
where 
\begin{equation}
\label{lb}
|\psi_{b|y}\rangle = \frac{( \id_{\mathcal A} \otimes \Pi_{b|y} ) |\psi\rangle}{\sqrt{\langle \psi | \id_{\mathcal A} \otimes \Pi_{b|y} | \psi \rangle}}.
\end{equation}

{\em Proof---}It is proven in \cite{Li:2023XXX} that $p(a,b|x,y)$ is a BPQS if and only if the zeros in $\{p(a,b|x,y)\}_{x \in X, y \in Y, a \in A, b\in B}$ do not admit a deterministic local hidden-variable model. This means that, for every assignment $f: (a|x) \cup (b|y) \rightarrow \{0,1\}$ satisfying $\sum_{a} f(a|x)=1$, $\forall x \in X$, and $\sum_{b} f(b|y)=1$, $\forall y \in Y$, there is a pair $\{(a|x),(b|y)\}$ for which $f(a|x)=f(b|y)=1$ and $p(a,b|x,y)=0$. 

Now notice that, for each $x \in X$, $s_x$, as defined in Theorem~1, is a set of mutually orthogonal states summing $\id_{\mathcal A}$.
Similarly, for each $y \in Y$, $s_y$ is a set of mutually orthogonal states summing $\id_{\mathcal A}$.

Second, notice that the element $p(a,b|x,y)$ is zero if and only if the corresponding $\rho^{\mathcal A}_{a|x}$ and $\rho^{\mathcal A}_{b|y}$, as defined in Theorem~1, are orthogonal.

Therefore, the impossibility of a deterministic local hidden-variable model
for the zeros in $\{p(a,b|x,y)\}_{x \in X, y \in Y, a \in A, b\in B}$ is equivalent to the impossibility of an assignment satisfying conditions (I') and (II') in Definition~2 for $S_{\mathcal A}$ and $S_{\mathcal B}$. Therefore, for every BPQS, following the prescription in Theorem~1, there are $S_{\mathcal A}$ and $S_{\mathcal B}$ that constitute a B-KS set and, therefore, a generalized KS set.

Theorem~1 can be used to prove an unexpected result. SI-C does not require a KS set, but only a SI-C set \cite{Yu:2012PRL,Bengtsson:2012PLA,Xu:2015PLA}. This might suggest that a BPQS can be produced by allowing the parties to measure a SI-C set that is not a KS set (as in \cite{Huang:2021PRL,Junior:2023PRR}). However,

{\em Corollary 1---}A BPQS cannot be produced by allowing the parties to measure a SI-C set that is not a KS set.

{\em Proof---}If the parties measure a SI-C set that is not a KS set, then Alice's postmeasurement states and Alice's premeasurement states conditioned to Bob's outcomes in Theorem~1 would admit a KS assignment. Therefore, the zeros of the correlation would admit a deterministic local hidden-variable model. However, a correlation is a BPQS if and only if its zeros do not admit a deterministic local hidden-variable model.

Theorem~1 also allows us to prove something that cannot be proven with any previous tool \cite{Li:2023XXX}.

{\em Theorem 2---}BPQSs are impossible with maximal observables in $|X|=|Y|=|A|=|B|=3$.

{\em Proof---}A BPQS with maximal observables in $|X|=|Y|=|A|=|B|=3$ would require the existence of a KS set in dimension $d=3$ with $3 \times 3 + 3 \times 3 = 18$ rank-one projectors. However, it has been proven that the smallest KS set in $d=3$ must have, at least, $24$ rank-one projectors \cite{Kirchweger:2023,Li:2024}. In fact, the smallest {\em known} KS set in $d=3$ has $31$ rank-one projectors \cite{Peres:1993}.

So far, all known examples of BPQSs require the initial state to be maximally entangled. Another interesting open problem \cite{Mancinska} is whether BPQSs {\em require} maximal entanglement. The fact that Theorem~1 does not assume that the state is maximally entangled suggests that maximal entanglement is not necessary, but further investigation is needed.

%%%%%%%%%%%%%%%%%%%%%%%%%%%%%%%%%%%%%%%%%%%%%%%%%%%%%%%%%%%%%%%%%%%

{\em BPQS with minimum resources---}Under the assumption that the players measure maximal observables, Theorem~1 allows us a different attack to Problems~1 and~2. We can consider all generalized KS sets (or, in the case of Problem~2, only the ones in $d=3$) with an increasingly larger number of elements and, for each of them, compute the way to distribute them between Alice and Bob (including the possibility of giving the same element to both of them) to generate a B-KS set $(S_{\mathcal A}, S_{\mathcal B})$ minimizing $|X| \times |Y|$. Then, among all of them, find the one that gives the BPQS minimizing $|X| \times |Y|$.

There is no comprehensive catalog of generalized KS sets in any dimension, but there is an extensive literature about KS and generalized KS sets with the smallest cardinalities known after $56$ years of research. In fact, it can be proven \cite{Xu:2022PRL} that the smallest KS set in any dimension requires $18$ rank-one projectors \cite{Cabello:1996PLA}. Applying the method described above to all the KS sets in \cite{Peres:1991JPA,Peres:1993,Kernaghan:1995PLA,Cabello:1996PLA,BubFOP1996,Bub:1997,Penrose:2000,LisonekPRA2014,Budroni:2022RMP} and all generalized KS sets in \cite{Toh:2013aCPL,Toh:2013bCPL}, we can formulate the following.

{\em Conjecture 1---}The BPQS with minimum $|X| \times |Y|$ is the following correlation: Alice and Bob share two ququarts in the state
\begin{equation}
|\psi \rangle = \frac{1}{2}\sum_{i=0}^3 |ii\rangle,
\end{equation}
Alice measures the three observables given by the following (unnormalized) orthogonal bases: 
\begin{subequations}
\begin{align}
&x=0 :{\scriptstyle \{(1,0,0,0),(0,1,0,0),(0,0,1,0),(0,0,0,1)\}}, \\
&x=1 : {\scriptstyle \{(1,1,1,1),(1,-1,1,-1),(1,1,-1,-1),(1,-1,-1,1)\}}, \\
&x=2 :{\scriptstyle \{(1,1,1,-1),(1,1,-1,1),(1,-1,1,1),(-1,1,1,1)\}},
\end{align}
\end{subequations}
and Bob measures the three observables given by the following (unnormalized) orthogonal bases: 
\begin{subequations}
\begin{align}
& y=0 : {\scriptstyle\{(1,1,0,0),(1,-1,0,0),(0,0,1,1),(0,0,1,-1)\}}, \\
& y=1 : {\scriptstyle\{(1,0,1,0),(0,1,0,1),(1,0,-1,0),(0,1,0,-1)\}}, \\
& y=2 : {\scriptstyle \{(1,0,0,1),(1,0,0,-1),(0,1,1,0),(0,1,-1,0)\}}.
\end{align}
\end{subequations}

%%%%%%%%%%%%%%%%%%%%%%%%%%%%%%%%%%%%%%%%%%%%%%%%%%%%%%%%%%%%%%%%%%%
% Fig. 1
%%%%%%%%%%%%%%%%%%%%%%%%%%%%%%%%%%%%%%%%%%%%%%%%%%%%%%%%%%%%%%%%%%%

\begin{figure}[t!]
\includegraphics[width=8cm]{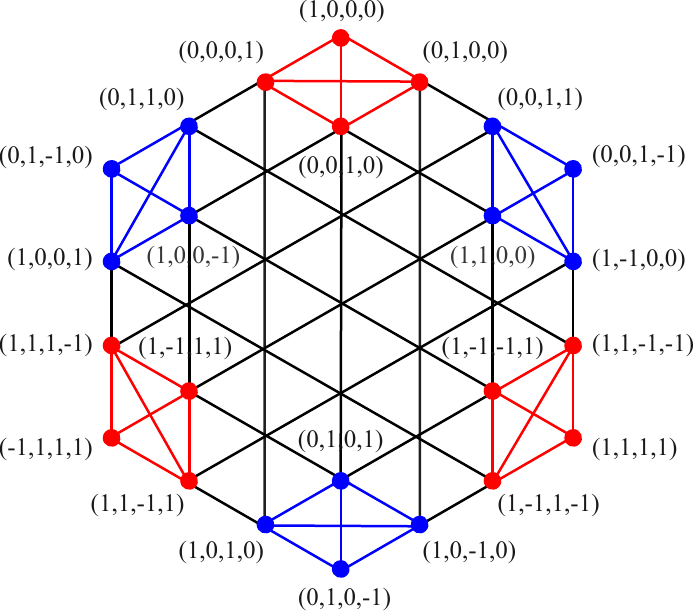}
\caption{Graph of orthogonality of the B-KS set leading to the BPQS of minimum $|X| \times |Y|$ with the method described in the text. Each dot represents a projector onto the indicated (unnormalized) state. Dots in the same line are orthogonal. This B-KS set is obtained by distributing between Alice and Bob the $24$ rank-one projectors of the KS set in $d=4$ in \cite{Peres:1991JPA}. $S_{\mathcal A}$ is defined by the $12$ red dots and defines $|X|=3$ observables with $|A|=4$ outcomes represented by the $4$ cliques of size $4$ in red. $S_{\mathcal B}$ is defined by the $12$ blue dots and defines $|Y|=3$ observables with $|B|=4$ outcomes represented by the $4$ cliques of size $4$ in blue. The underlying (uncolored and without vectors) figure is taken from \cite{Blanchfield:2014}.
}
\label{Fig1}
\end{figure}

%%%%%%%%%%%%%%%%%%%%%%%%%%%%%%%%%%%%%%%%%%%%%%%%%%%%%%%%%%%%%%%%%%%

{\em Evidence 1---}The B-KS set leading to the BPQS of minimum $|X| \times |Y|$ found with the method described before is the one shown in Fig.~\ref{Fig1}.

The correlation in Conjecture~1 is not new. It is the ``magic-square'' correlation introduced in \cite{Cabello:2001PRLa,Cabello:2001PRLb}, which has been used extensively in the literature (e.g., \cite{Aravind:2004AJP,GBT05,Bravyi:2018SCI,Ji:2021CACM}) and has been experimentally tested with hyperentangled photons \cite{CinelliPRL2005,YangPRL2005,Aolita:2012PRA,Xu:2022PRL}.

{\em Conjecture 2---}The qutrit-qutrit BPQS with minimum $|X| \times |Y|$ is the following correlation: Alice and Bob share the state
\begin{equation}
\label{state}
|\psi\rangle = \frac{1}{\sqrt{3}} \sum_{i=0}^2 |ii\rangle,
\end{equation}
Alice measures the nine observables given by the following (unnormalized) orthogonal bases:
\begin{subequations}
\begin{align}
&x=0 :\{(1,0,0),(0, 1, 1),(0, 1, -1)\}, \\
&x=1 :\{(0, 1, 0),(1, 0, 1),(1, 0, -1)\}, \\
&x=2 :\{(0, 0, 1),(1, 1, 0),(1, -1, 0)\}, \\
&x=3 :\{(1, 0, 0),(0, 1, \sqrt{2}),(0, \sqrt{2}, -1)\}, \\
&x=4 :\{(1, 0, 0),(0, 1, -\sqrt{2}),(0, \sqrt{2}, 1)\}, \\
&x=5 :\{(0, 1, 0),(1, 0, \sqrt{2}),(\sqrt{2}, 0, -1)\}, \\
&x=6 :\{(0, 1, 0),(1, 0, -\sqrt{2}),(\sqrt{2}, 0, 1)\}, \\
&x=7 :\{(0, 0, 1),(1, \sqrt{2}, 0),(\sqrt{2}, -1, 0)\}, \\
&x=8 :\{(0, 0, 1),(1, -\sqrt{2}, 0),(\sqrt{2}, 1, 0)\},
\end{align}
\end{subequations}
and Bob measures the seven observables given by the following (unnormalized) orthogonal bases:
\begin{subequations}
\begin{align}
&y=0 :\{(1, 0, 0),(0, 1, 0),(0, 0, 1)\}, \\
&y=1 :\{(1, 1, 0),(1, -1, \sqrt{2}),(-1, 1, \sqrt{2})\}, \\
&y=2 :\{(1, -1, 0),(1, 1, \sqrt{2}),(1, 1, -\sqrt{2})\}, \\
&y=3 :\{(1, 0, 1),(1, \sqrt{2}, -1),(-1, \sqrt{2}, 1)\}, \\
&y=4 :\{(1, 0, -1),(1, \sqrt{2}, 1),(1, -\sqrt{2}, 1)\}, \\
&y=5 :\{(0, 1, 1),(\sqrt{2}, 1, -1),(\sqrt{2}, -1, 1) \}, \\
&y=6 :\{(0, 1, -1),(\sqrt{2}, 1, 1),(-\sqrt{2}, 1, 1) \}.
\end{align}
\end{subequations}

%%%%%%%%%%%%%%%%%%%%%%%%%%%%%%%%%%%%%%%%%%%%%%%%%%%%%%%%%%%%%%%%%%%
% Fig. 2
%%%%%%%%%%%%%%%%%%%%%%%%%%%%%%%%%%%%%%%%%%%%%%%%%%%%%%%%%%%%%%%%%%%

\begin{figure}[t!]
\includegraphics[width=8.0cm]{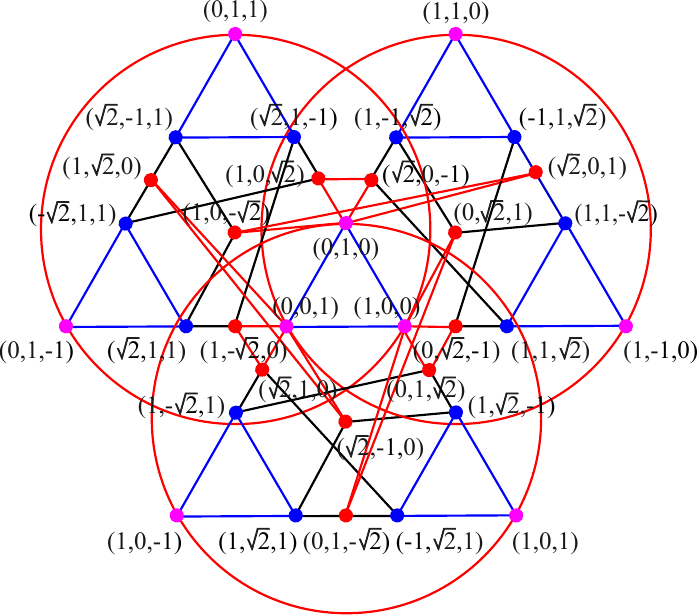}
\caption{Graph of orthogonality of the B-KS set leading to the qutrit-qutrit BPQS of minimum $|X| \times |Y|$ with the method described in the text. Each dot represents a projector onto the indicated (unnormalized) state. Adjacent dots are orthogonal. This B-KS is obtained by distributing between Alice and Bob the $33$ rank-one projectors of the KS set in $d=3$ in \cite{Peres:1991JPA} (but one could use any other KS set of the Peres-Penrose family \cite{Penrose:2000,gould2010isomorphism,bengtsson2012gleason}) in a particular way. Notice that, unlike in the example of Fig.~\ref{Fig1}, here the same rank-one projector is sometimes given to both parties. $S_{\mathcal A}$ is defined by the $16$ red dots and the $9$ violet dots and defines $|X|=9$ observables with $|A|=3$ outcomes (represented by circumferences and triangles in red). $S_{\mathcal B}$ is defined by the $12$ blue dots and the $9$ violet dots and defines $|Y|=7$ observables with $|B|=3$ outcomes (represented by triangles in blue). The underlying (uncolored and without vectors) figure is taken from \cite{Blanchfield:2014}.} 
\label{Fig2}
\end{figure}

%%%%%%%%%%%%%%%%%%%%%%%%%%%%%%%%%%%%%%%%%%%%%%%%%%%%%%%%%%%%%%%%%%%

{\em Evidence 2---}The B-KS set leading to the BPQS of minimum $|X| \times |Y|$ found with the method described is the one shown in Fig.~\ref{Fig2}.

The correlation in Conjecture~2 substantially improves the qutrit-qutrit correlation in the literature with the smallest $|X| \times |Y|$, namely, $|X|=31$, $|Y|=17$, $|A|=2$, $|B|=3$ \cite{Elby:1992PLA,Renner2004b,Peres:1993}.

%%%%%%%%%%%%%%%%%%%%%%%%%%%%%%%%%%%%%%%%%%%%%%%%%%%%%%%%%%%%%%%%%%%

{\em BPQS and SI-C---}Theorem~1 solves Problem~3. A BPQS {\em requires} a KS set because every BPQS defines a KS set. This gives us a surprising lesson: while KS sets were initially important because of SI-C \cite{Cabello:2008PRL,Badziag:2009PRL,Kirchmair:2009NAT,Amselem:2009PRL}, now we know that they are not needed for SI-C \cite{Yu:2012PRL,Bengtsson:2012PLA,Xu:2015PLA}. However, Theorem~1 shows that KS sets are {\em necessary} (and, with maximal entanglement, sufficient) for BPQSs. That is, KS sets are fundamental in {\em Bell nonlocality} and not so fundamental in SI-C. More generally, Theorem~1 also tells us that KS sets are crucial for groundbreaking results in quantum computation, information, and foundations [see, e.g., second, third, and fifth results in the introduction of this Letter]. Hopefully, the results and conjectures presented help to stimulate even further interest in perfect quantum correlations and KS sets.
\\

{\em Acknowledgments---}I thank I.\ Bengtsson, K.\ Bharti, J.\ R.\ Gonzales-Ureta, L.\ Porto, R.\ Ramanathan, D.\ Saha, S.\ Trandafir, Z.-P.\ Xu, and E.\ Zambrini Cruzeiro for helpful discussions and comments.
This work was supported by the EU-funded project \href{10.3030/101070558}{FoQaCiA}, the \href{10.13039/501100011033}{MCINN/AEI} (Project No.\ PID2020-113738GB-I00), and the Wallenberg Center for Quantum Technology (WACQT).

%%%%%%%%%%%%%%%%%%%%%%%%%%%%%%%%%%%%%%%%%%%%%%%%%%%%%%%%%%%%%%%%%%%

%\bibliographystyle{apsrev4-2} %Remove to allow the longbibliography option to work
%\bibliography{common}

%

%%%%%%%%%%%%%%%%%%%%%%%%%%%%%%%%%%%%%%%%%%%%%%%%%%%%%%%%%%%%%%%%%%%

\end{document}